\documentclass[11pt]{article}
\usepackage[english]{babel}
\usepackage[a4paper,top=2cm,bottom=2cm,left=2.5cm,right=2.5cm,marginparwidth=1.75cm]{geometry}
\usepackage{setspace}
\usepackage{enumitem}
\setlist[description,1]{leftmargin=0em,labelindent=*}
\usepackage{amsmath}
\usepackage{amssymb}
\usepackage{graphicx}
\usepackage[colorlinks=true, allcolors=blue]{hyperref}
\usepackage{authblk}
\usepackage{xcolor} 
\definecolor{teal}{rgb}{0,0.63,0.67}
\definecolor{darkteal}{rgb}{0, 0.44, 0.47}

\usepackage[most]{tcolorbox} % Load the tcolorbox package

\makeatletter

\renewcommand\AB@affilsepx{, }

\title{Why the Northern Hemisphere Needs a 30–40 m Telescope and the Science at Stake: A Low Surface Brightness Science Case}

\author[1]{Mireia Montes}
\author[2,3]{Ignacio Trujillo}
\author[4]{David Mart\'inez Delgado}
\author[4]{Borja Anguiano}
\author[5]{Magda Arnaboldi} 
\author[2,3]{Michael A. Beasley}
\author[6,7]{Fernando Buitrago}
\author[8]{Michele Cantiello}
\author[9]{Andr\'es del Pino}
\author[10]{Amandine Doliva-Dolinsky} 
\author[11]{Helena Dom\'inguez-S\'anchez}
\author[12]{Mauro D'Onofrio}
\author[13]{Pierre-Alain Duc} 
\author[14]{Katja Fahrion} 
\author[2,3]{Anna Ferr\'e-Mateu}
\author[2,3]{Carme Gallart}
\author[15]{Nina Hatch} 
\author[16]{Enrica Iodice}
\author[9]{Yolanda Jim\'enez-Teja} 
\author[17]{Francine Marleau}
\author[18]{Chris Mihos}
\author[19]{Nicola Napolitano}
\author[20]{Agnieszka Pollo}
\author[21]{Javier Rom\'an}
\author[8]{Joanna Sakowska} 
\author[2,3]{Jorge S\'anchez Almeida}
\author[21]{Patricia S\'anchez-Bl\'azquez} 
\author[16]{Marilena Spavone}
\author[2,3]{Guillaume Thomas}
\author[2,3]{Eva Villaver}

\affil[1]{Institute of Space Sciences (ICE, CSIC), Spain}
\affil[2]{Instituto de Astrofísica de Canarias (IAC), Spain}
\affil[3]{Universidad de La Laguna (ULL), Spain}
\affil[4]{Centro de Estudios de Física del Cosmos de Arag\'on (CEFCA), Spain}
\affil[5]{European Southern Observatory (ESO), Germany}
\affil[6]{Universidad de Valladolid (UVa), Spain}
\affil[7]{Instituto de Astrofísica e Ciências do Espaço (Portugal)}
\affil[8]{INAF-Astronomical Observatory of Teramo, Italy}
\affil[9]{Instituto de Astrof\'isica de Andaluc\'ia (IAA), Spain}

\affil[10]{University of Surrey, UK}
\affil[11]{Instituto de F\'isica de Cantabria (IFCA), Spain}
\affil[12]{University of Padova, Italy}
\affil[13]{Observatory of Strasbourg, France}
\affil[14]{University of Viena, Austria}
\affil[15]{University of Notthingham, UK}
\affil[16]{INAF-Astronomical Observatory of Capodimonte, Italy}
\affil[17]{Universit\"at Innsbruck, Austria}

\affil[18]{Case Western Reverse University, US}
\affil[19]{University Federico II Naples, Italy}
\affil[20]{National Centre For Nuclear Research, Poland}

\affil[21]{Universidad Complutense de Madrid (UCM), Spain}

\date{}
\begin{document}
\maketitle

\newpage
\vspace{-0.2cm}
\begin{tcolorbox}[
    colback=teal!5!white, % Background color (20% of myblue, 80% white)
    colframe=teal,         % Frame color
    arc=2mm,                 % Rounded corner radius
    boxsep=0.6mm,              % Space between box and content
    left=2mm, right=2mm, top=2mm, bottom=2mm % Padding inside the box
]
The Extragalactic Low Surface Brightness (LSB, $\mu_V\gtrsim 27$ mag/arcsec$^2$) Universe represents a crucial, yet largely unseen, frontier in modern astrophysics. This faint realm holds the keys to completing our understanding of galaxy evolution, hierarchical assembly, and even the fundamental nature of dark matter. Our current theoretical models are inherently incomplete, largely mirroring the properties of the brightest, most easily observed objects. To overcome this critical bias and unlock the secrets of this realm, a transformative leap in observational capability is required. A 30 to 40m class telescope, leveraging unprecedented sensitivity and spatial resolution, especially with adaptive optics, is the essential tool to fundamentally probe these faint, low-density stellar regimes. This white paper details the transformative LSB science that such a facility, strategically positioned in the Northern Hemisphere (NH) to access crucial nearby structures and rich environments, can achieve.
\end{tcolorbox}

\vspace{-0.6cm}
\section{General Cases}
\vspace{-0.2cm}
A 30m class telescope will revolutionize our understanding of the LSB Universe in two main ways. First, its high-resolution imaging, aided by adaptive optics (AO), will enable us to resolve individual stars in nearby LSB systems (out to $\sim 15$ Mpc) [1]. This capability is essential for dissecting stellar populations, allowing us to obtain detailed chemical compositions and accurate stellar ages. Second, the telescope's immense sensitivity will be crucial for analysing the integrated light of more distant systems ($>15$ Mpc), providing key constraints on their overall stellar populations and formation histories. 

\vspace{-0.3cm}
\subsubsection*{\underline{The building blocks of the Universe}}
\vspace{-0.2cm}
LSB galaxies and dwarf systems constitute the bulk of the galaxy population by number [2] and are fundamental to understanding cosmic structure. They are crucial for two primary reasons: first, they are the key to understanding star formation efficiency in low-mass dark matter halos [3]; and second, they are the recognized building blocks through which larger galaxies, including the Milky Way, are assembled [4, 5]. Despite their cosmological significance, these galaxies, which represent the faintest and most dark-matter-dominated systems in the Universe, remain poorly characterized. Yet, analyzing their star formation histories provides one of the most stringent tests of contemporary galaxy formation models, holding the key to unraveling the complex interplay of physics driving galaxy evolution.

\noindent A 30m class telescope will finally provide statistically significant samples and probe significantly lower-luminosity galaxies than currently possible. This leap in capability will clarify the formation mechanisms of these galaxies, test the influence of feedback processes (such as supernovae) in low-mass dark matter halos, determine the mechanism for their quenching, and constrain how cosmic reionization shaped their early evolution [6,7]. 

\vspace{-0.3cm}

\subsubsection*{\underline{The history of assembly of structure}}
\vspace{-0.2cm}
The most revealing signature of the hierarchical assembly in galaxies, groups and clusters of galaxies is the intricate web of faint stellar features that surrounds them. These outer regions constitute a major component of the host's halo, preserving a detailed record of galaxies destroyed during the hierarchical assembly of the host. They, thus, provide direct evidence of the accretion history of the host [8, 9]. These regions are composed of a globally smooth component such as the stellar halo and the intragroup/intracluster light (intrahalo light or IHL for short) [10, 11], but also contain substructures formed by the disruption of satellite galaxies, including stellar streams, shells, and other tidal signatures. A 30m class telescope will be instrumental for characterizing the stellar populations of these outer regions.

\noindent For individual galaxies, this capability will allow us to reconstruct their entire formation history across a statistically large and diverse sample [12]. This crucial step extends our view beyond the Milky Way and M31, providing the necessary dataset to robustly compare against cosmological simulations, which is an essential exercise given the inherent stochastic nature of galaxy assembly. Furthermore, a 30m telescope will also enable ``chemical tagging" to trace and discover faint streams, analogous to the techniques employed in the Milky Way halo [13]. It will also reveal the location of galaxy edges or truncations—both in nearby systems via resolved star counts and in more distant and fainter systems via integrated light—providing key boundary conditions for models of galaxy assembly and stellar migration [14].
For galaxy groups and clusters, it will allow us to unveil their recent dynamical history via the study of tidal features and substructure within the IHL [15, 16]. Furthermore, for the closest clusters, the telescope's deep sensitivity will enable spatially resolved mapping of the stellar populations across the cluster (see Section 2 for more details).

\vspace{-0.4cm}
\subsubsection*{\underline{The nature of dark matter}} 
\vspace{-0.1cm}
The stars of the LSB Universe exist in regimes where dark matter dominates the gravitational potential. This makes these stars a unique, collisionless tracer of the dark matter halos they inhabit. Without the precise kinematic information (velocities and velocity dispersions) made accessible only by a 30m class telescope, stellar orbits cannot be accurately measured. Consequently, the gravitational potential cannot be constrained with the precision needed to differentiate between competing models of dark matter. Accessing this kinematic data is therefore essential for testing the fundamental nature of dark matter itself.
\vspace{-0.1cm}
\begin{description} 

\item[LSB Galaxies:] The capabilities of a 30m telescope will allow measuring the stellar velocity dispersions and rotation curves to probe the dark matter density profiles of LSB galaxies. Particularly interesting are LSB systems with low stellar masses, specifically those below 10$^5$–10$^6$ M$_\odot$. In these low-mass halos, stellar feedback is expected to have a negligible effect on the central dark matter distribution [17, 18]. This makes the internal stellar kinematics of these systems critical tracers of the central gravitational potential, allowing for the direct discrimination between ``core" and ``cusp" halo profiles.
Probing these dark matter density profiles is critical for testing the fundamental nature of dark matter itself [19, 20, 21]. Furthermore, the 30m telescope will enable an independent determination of the LSB galaxies' dark matter content through analysis of the dynamics of their associated globular cluster systems [22].

\item[Extragalactic Stellar Streams:] Obtaining kinematic measurements of stellar streams, either through individual stars [23] or via spectroscopy of tracers like Planetary Nebulae (PNe) or globular clusters (GCs) [24], is critical for breaking the well-known degeneracies found in N-body simulations [25], thereby enabling definitive constraints on the total mass, shape, and gravitational potential of the dark matter halo of their host galaxies.
\item[Intrahalo light:] Similarly as in stellar streams, the stars in the smooth IHL are an effective, large-scale tracer of the global gravitational potential of its host halo [26]. Crucially, the total stellar mass and spatial distribution of these halos are a direct consequence of the nature of dark matter itself [27]. Therefore, acquiring kinematic measurements of IHL tracers, individual stars, GCs [28] and PNe [29] or integrated light spectroscopy when there are few tracers, is essential. These measurements are vital to map the host halo's global potential, probe its properties, identify substructure, and ultimately place rigorous constraints on the nature of dark matter.
\item[Detecting Dark Subhalos:] These are predicted cold dark matter halos with masses smaller than $10^9$ M$_\odot$ that have never formed stars. The passage of these dark matter sub-haloes across other visible structures (tidal streams, gravitational lenses) will leave imprints that will be observable and quantifiable [30] beyond the Local Group with a 30m telescope. However, they may also be detected through a characteristic ring-like emission pattern to be expected in H$\alpha$, caused by the recombination of the outer shell of the gas in the halos, ionized by the cosmic UV background [31]. These very faint signals need a 30m telescope to be detected.
\end{description}

\noindent $\bullet$ In addition, a 30m telescope will allow for follow-up of peculiar objects discovered in ultra-deep imaging surveys like LIGHTS, Euclid, ARRAKIHS and part of the LSST footprint. 

\vspace{-0.3cm}
\section{Objects in the Northern Hemisphere}
\vspace{-0.2cm}
\begin{description}
    
\item[\textcolor{teal}{M31 and its Satellites:}] 
A 30m class telescope will provide unprecedented insight into the evolutionary history of M31. Using AO, it will measure the proper motions, radial velocities and chemistry of individual solar-type stars in the outer halo, which together with deep color-magnitude diagrams, will enable the reconstruction of M31's star formation and chemical enrichment histories. Furthermore, the telescope's sensitivity is essential for studying M31's large satellite population [33, 34]. Specifically, the faintest satellites (below M$_*\sim$10$^5$–10$^6$ M$_\odot$) [35, 36] which are critical for constraining competing dark matter models. It will also confirm candidate ultra-faint dwarfs, completing the galaxy's luminosity function. The study of the low surface brightness components of the M31 systems is highly complementary to the science proposed in the white paper by Gallart et al.

\vspace{-0.2cm}
\item[\textcolor{teal}{The Richest Galaxy Clusters:}] 
The NH hosts the richest and most relaxed galaxy clusters in the nearby universe: Coma (z=0.023, 100 Mpc) and Perseus (z=0.018, 75 Mpc). These massive overdensities are fertile grounds for discovering and studying the most extreme systems. Their dense environments are ideal for the formation —or survival— of galaxies exhibiting either very little or an unusually large amount of dark matter. Recent discoveries of ultradiffuse galaxies exemplify this diversity [37], as these very faint systems occur in large numbers within rich clusters, with increasingly extreme examples continuing to be found. Due to their extreme faintness, these systems are accessible only through the deep spectroscopy afforded by instruments on 30m class telescopes.\\
These massive clusters also serve as perfect laboratories for the in-depth study of their IHL. A 30m telescope will extend spectroscopy into the ICL's faintest limits [38], providing the necessary signal-to-noise to characterize the stellar populations of the diffuse light, which is essential for studying assembly processes throughout the cluster. Furthermore, kinematic tracers like GCs and PNe will allow us to obtain detailed kinematics and map the gravitational potential of these massive clusters.
\vspace{-0.2cm}
\item[\textcolor{teal}{Nearby Milky-Way Analogs:}] $\sim$2/3 excess of the largest galaxies in the sky (diameter $> 10$') are observable from the NH. That allows the characterization in detail of their satellite population and their outer regions. In particular, the three nearest edge-on Milky Way analogs are in the NH (NGC 891, NGC 4565 and NGC 5907 at $\sim$10 Mpc) which are exceptional laboratories for studying vertical disk structure, disk stability in the outskirts and faint stellar halos. Their edge-on orientation allows exploration of vertical disk profiles and of stellar migration and heating due to bombardment by dark-matter substructures [39, 40]. Questions such as whether the Milky Way’s inner halo is typical and how disk thickness evolves with height above the plane can be addressed in these galaxies with the power of a 30m-class facility.
\vspace{-0.2cm}
\item[\textcolor{teal}{The Virgo Cluster:}] This nearby cluster is composed of the main galaxy groups (Virgo A and B) and numerous smaller substructures currently accreting along cosmic filaments [41, 42].
Its proximity and varied environments —dense cores to low-density infall regions— make it a critical laboratory for understanding LSB galaxies and their stellar properties [43, 44]. A 30m class telescope is also uniquely positioned to conduct detailed, resolved studies of diffuse light across this full range of environments.

\end{description}
\vspace{-0.85cm}
\section{Other considerations }
\vspace{-0.3cm}
\noindent The level of detail provided by a 30m class telescope is necessary to understand the full history and composition of our cosmic landscape. The light collecting power and spacial resolution of such a facility are required to advance this field, which is currently limited to studying only a handful of objects or relying primarily on shallow broadband imaging. For the LSB science described throughout this paper, a field of view larger than what is typical in the currently planned 30-m telescopes would substantially enhance the efficiency and statistical power.
\vspace{-0.2cm}

\subsection*{References}
[1] Skidmore et al. 2015, RAA, 15, 1945 [2] Baldry et al. 2008, MNRAS, 388, 945 $\bullet$ [3] Read et al. 2017, MNRAS, 467, 2019 $\bullet$ [4] Searle \& Zinn 1978, ApJ, 225, 357 $\bullet$ [5] Moore et al. 1999, ApJ, 524, 19 $\bullet$ [6] Benson et al. 2002, MNRAS, 333, 117 $\bullet$ [7] Fitts et al. 2017, MNRAS, 471, 3547 
$\bullet$ [8] McConnachie et al. 2009, Nature, 461, 66 
$\bullet$ [9] Mihos et al. 2005, ApJ, 631, 41 
$\bullet$ [10] Bullock \& Johnston 2005, ApJ, 635, 931 
$\bullet$ [11] Monachesi et al. 2016, MNRAS, 457, 1419 
$\bullet$ [12] Mart\'inez-Delgado et al. 2025, arXiv:2509.14038 
$\bullet$ [13] Smercina et al. 2020, ApJ, 905, 60
$\bullet$ [14] Buitrago \& Trujillo, 2024, A\&A, 682, 110
$\bullet$ [15] Jim\'enez-Teja et al. 2025, A\&A, 694, 216
$\bullet$ [16] Ellien et al. 2025, A\&A, 698, 134
$\bullet$ [17] Governato et al. 2012, MNRAS, 422, 1231
$\bullet$ [18] Di Cintio et al. 2014, MNRAS, 441, 2986
$\bullet$ [19] de Blok, AdAst, 2010, 5
$\bullet$ [20] Pontzen \& Governato, 2014, Nature, 506, 171
$\bullet$ [21] S\'anchez Almeida et al. 2024, ApJ, 973, 15
$\bullet$ [22] Beasley et al. 2016, ApJ, 819, 20
$\bullet$ [23] Toloba et al. 2016, ApJ, 824, 35
$\bullet$ [24] Foster et al. 2014, MNRAS, 442, 3544
$\bullet$ [25] Walder et al. 2025, ApJ, 994, 36
$\bullet$ [26] Montes \& Trujillo, 2019, MNRAS, 482, 2838
$\bullet$ [27] Boehm et al. 2014, MNRAS, 445, 31
$\bullet$ [28] Reina-Campos et al. 2022, MNRAS, 513, 3925
$\bullet$ [29] Longobardi et al. 2018, A\&A, 620, 111
$\bullet$ [30] Nguyen et al. 2025, arXiv:2512.07960
$\bullet$ [31] Sykes et al. 2019, MNRAS, 487, 609
$\bullet$ [32] Gallart et al. 2005, ARA\&A, 43, 387
$\bullet$ [33] McConnachie 2012, AJ, 144, 4
$\bullet$ [34] Doliva-Dolinsky et al. 2023, ApJ, 952, 72
$\bullet$ [35] Collins et al. 2011. MNRAS, 417, 1170
$\bullet$ [36] Kirby et al. 2020, AJ, 159, 46
$\bullet$ [37] van Dokkum et al. 2015, ApJ, 798, 45
$\bullet$ [38] Gu et al. 2020, ApJ, 894, 32
$\bullet$ [39] Villalobos \& Helmi 2008, MNRAS, 391, 1806
$\bullet$ [40] Minchev et al. 2012, A\&A, 548, A127
$\bullet$ [41] Gavazzi, et al. 1999, MNRAS, 304, 595
$\bullet$ [42] Cantiello, et al. 2024, ApJ, 966, 145
$\bullet$ [43] Mihos et al. 2015, ApJ, 809, 21
$\bullet$ [44] Ferr\'e-Mateu et al. 2023, MNRAS, 526, 4735

\end{document}